# E-Governance: Past, Present and Future in India

Nikita Yadav
Research Scholar, Singhania University,
Pacheri Bari, Rajasthan

V. B. Singh
Delhi College of Arts & Commerce, University of Delhi, Delhi

## ABSTRACT
Online working of a government or providing its services online to its citizens at their door step is known as E-Governance. E-Governance is E-Commerce technology means online availability of government services. The technology and the methods used in E-Governance project provide a roadmap for efficient delivery of services at the door step. In today's time the development of any country depends on the uses of E-Governance and also their penetration. Development of any country can be judge by the scope of E-Governance in that country. Moreover, today's government has also full faith in E-Governance and its widespread network across the world proves it. Due to widespread demand of E-governance and exponentially increasing size of data, new technologies like Open source solutions and cloud computing need to be incorporated. In this paper, the latest trends of technology that the government of most of the country has adopted have been discussed. While working on this project we have concluded that E-Governance has made the working of government more efficient and more transparent to its citizens We have also presented an exhaustive list of E-Governance projects which is currently being used in India and in international scenario. We have provided a mechanism for improving E-Governance by including technologies such as Open Source and Cloud Computing.

### General Terms
Information Technology

## 1. INTRODUCTION
Today's is the world of change and revolution. Things which are not dynamic are considered as mortal. The rise of e-government has been one of the most striking developments of the web Global shifts towards increased deployment of IT by governments emerged in the nineties, with the advent of the World Wide Web. The technology as well as e-governance initiatives have come a long way since then. Recognizing the increasing importance of electronics, the Government of India established the Department of Electronics in 1970.[7] The subsequent establishment of the National Informatics centre (NIc) in 1977 was the first major step towards e-Governance in India as it brought 'information' and its communication in focus. However, the main thrust for e-Governance was provided by the launching of NIcNET in 1987 – the national satellite-based computer network. This was followed by the launch of the District Information System of the National Informatics centre (DISNIc) program to computerize all district offices in the Country for which free hardware and software was offered to the State Governments. NIcNET was extended via the State capitals to all district headquarters by 1990 [7].

E-Governance was started in India by AHSHAYA in Kerala. This project involves setting up around 5000 multipurpose community technology centers called Akshaya e- Kendra's across Kerala. Run by private entrepreneurs, each e-Kendra set up within 2-3 kilometers of every household, will cater to the requirements of around 1000-3000 families to make available the power of networking and connectivity to common man. Akshaya is a social and economic catalyst focusing on the various facets of e-learning, e-transaction, e-governance, information and communication.

E-Governance is not only popular in India but also worldwide. To make working of government more efficient, responsive and transparent many developed and developing countries have taken some useful steps for the expansion of e-governance in their respective countries. Countries which are in the race of e-governance implementation are UK, USA, New Zealand, Brazil etc. Below following countries with their respective projects has been described:-

**USA:** On July 18, 2001 Task Force to identify priority actions that achieve strategic improvements in government and set in motion a transformation of government around citizen needs was done [10].

Various projects:-

1. Recreation One Stop: - This project was started on 31 April 2002. It aimed to:-
- Agreement with private sector reached on implementation of new recreation online projects.
- Additional recreation projects available online.

2. EZ Tax filling:-This project was started on 31april, 2002. It aimed to:-

- Filling and refunding of taxes online.
- Initial deployment of industry partnership free e-filling solution for 2003season.

3. Federal Assets Sales:-started on 31march, 2003. It aimed to:-

- Develop pilot business integration
- Re-host federal sales.

4. E-Payroll/HR: - started on 31 March, 2002. Aimed to:-

- complete and submit business case to PMC
- integrated enterprise architecture

5. E-Authentication: - started on 1july, 2002. Aimed at:-

- Initial authentication gateway prototype
- Full deployment

**UK:** In April 2000, the cabinet Office in the UK came out with the document 'E-Government: A Strategic Framework for Public Services in the Information Age' [9].

Various projects:-

1. Cornwall Electronic Health Record Pilot: - this project was started in April 2000[11]. It aimed to:-

- Pan-community EHR demonstrator
- Connect all General Practitioners to NHSnet (national-level NHS Virtual Private Network or intranet)
- 24 hour emergency care record
- Common information architecture
- Condition-specific care modules – mental health, coronary heart disease, diabetic care

2. Go-between Project: - this project is for Calne Community Area in Wiltshire. This project is designed to assist a local rural community organize its transport services by identifying unmet local demand and matching it with spare capacity [12]

3. Plymouth Bus Project:-Commissioned by Plymouth City Council to undertake a comprehensive study of the bus network and propose initiatives for its future development and investment. (Completed in January 2009) [12]

4. Integrating Transport:-Started in South West Hertfordshire. Lead consultant on a project commissioned by Watford Borough Council. Preparation of a sustainable transport strategy for implementation over the next ten years. (Completed October 2008)[12]

**New Zealand**

In May 2000,Realising the importance of opportunities offered by ICT, the New Zealand Government came out with its e-government vision document and an 'E-Government Unit' was established by the State Services commission

Various projects:-

The Treaty of Wetangi: It is the founding document of New Zealand. This website, launched in 2003, to provide information and resources for an informed understanding and greater public knowledge of the treaty[13].

As people's want advancement in their lifestyle same they want in their government. Success or failure of any government depends on three things:

- The way the government works.
- Transparency in its working and information.
- Communication to its citizen.

In developing countries, Government beneficiary scheme can be reached to the lowest pyramid of society efficiently by implementing E-Governance projects. It can bridge the gap between rich and poor. Rest of the paper is organized as follows. Section 1.1 and 1.2 of the paper describes about E-Governance. Applications of E-Governance and challenges in implementing E-Governance have been described in section 2, 3, 4 and 5.In section 6; future technologies for e-governance are described.

**1.1 E-Governance:** Use of internet by the government to provide its services at the door step of customers, business and other stakeholder.

In E-Governance, government makes best possible use of internet technology to communicate and provide information to common peoples and businessman. Today, electricity, water, phone and all kinds of bills can be paid over the internet. All this is what government and citizens is using and doing. All are dependent on internet and when citizens depends on government internet services all that come is E-Governance.

There are four pillars of E-Governance:-

1. CONNECTIVITY:-Connectivity is required to connect the people to the services of the government. There should be a strong connectivity for an effective e-governance.
2. KNOWLEDGE: - Here knowledge refers to IT knowledge. Government should employ skill full engineers who can handle the e-governance in an efficient way. These engineers also handle all kind of fault that may occur during the working of e-governance.
3. DATA CONTENT: - To share any kind of knowledge or information over the internet, there should be its database. This database should have the data content which is related to government services.
4. CAPITAL:-Capital can be on public or private partnership. It refers to money used by government to provide their services or to that sector of the economy based on its operation

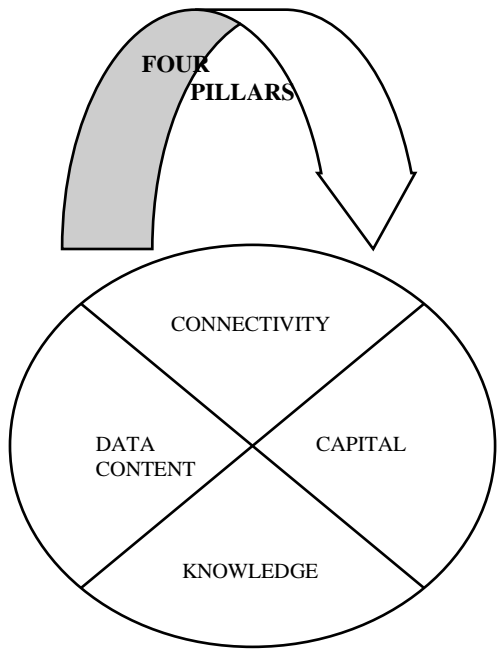

**Fig 1: pillars of e-governance**

## 1.1 E-governance models

E-GOVERNANCE services can be shared between citizens, businessman, government and employees. These four models of e-governance are as:-

1. Government to citizens (G2C)
2. Government to government (G2G)
3. Government to employees (G2E)
4. Government to businessman (G2B)

**Government to citizens (G2C):-**This model of e-governance refers to the government services which are shared by citizens. Here, citizens visit to the link of services that they want to use. This models strong the bond between government and its citizen. Type of services which are provided by this model includes:-

- Payment of online bills such as electricity, water, telephone bills etc.
- Online registration of applications.
- Copies of land-record.
- Online filling of complaints.
- Availability of any kind of online information.

**Government to government (G2G):-**This model refers to the services which are shared between the governments. There is lots of information that need to be shared between various government agencies, department and organizations. These types of services or information are as:-

- Sharing of information between police department of various state.
- Government document exchange which includes preparation, approval, distribution, and storage of all governmental documents is also done through e-governance.

- Most of the finance and budget work are also done through e-governance.

**Government to businessmen (G2B):-**Through this model, bond between private sector and government increase and businessmen use to communicate. They share information through this model like:-

- Collection of taxes.
- Rejection and approval of patent is also done by this model.
- Payment of all kind of bills and penalty.
- Sharing of all kind of information, rules and data.
- Complaints or any kind of dissatisfaction can be shown by this.

**Government to employees (G2E):-**This model increases the transparency between government and its employee. Here, employee can keeps a check on the functioning and working of government and government can keeps on its employees. Information that can be shared by this model:-

- All kind of data submission(attendance record, employee record etc) from various government offices is done by this model
- Employee can file all kinds of complaints and dissatisfaction by this model.
- All kind of rule- regulation and information for employees can be shared by this.
- Employees can check their payment and working record.
- Employees can register all kind of working forms online.

## 2 DIFFERENT AREAS OF E-GOVERNANCE

Today area of e-governance is very wide. E-Governance is implemented by government in almost every field. From urban states to rural areas and from politics to teaching- Governance has spread its root everywhere. Either its public or private sector, common man or businessman all is largely dependent on e-governance. Here we have presented different areas where e-governance is widely used. In the following section, we are describing the projects used in urban and rural areas of India.

## 2.1 E-Governance projects in urban areas:-

*2.1.1Transportation:-* Services provided by e-governance in this area are:-

- Issuance of Time Table of buses.
- Provision of booking facility for Interstate transport.
- Transportation Improvement Program.
- Regional Transportation Plans.
- Congestion Management Process.
- Transportation Demand Management.

Various projects [1]:-

1. CFST:-Citizen Friendly Services of Transport Department by Andhra Pradesh government to provide services such as Issue of learner licensees

,Issue of driving licenses ,Renewal of driving licenses etc
2. Vahan and Sarathi:-The backend applications Vahan&Sarathi help in speeding the overall work flow in the transport department but Tamil Nadu govt.
3. OSRTC:- The Orissa State Road Transport Corporation project was started to provide transport related facilities online
4. HRTC: - Himachal Road Transport Corporation project is for online bookings, cancellation of seats, for enquiry about departure of buses, availability of seats and buses etc.

*2.1.2 Online payment of bills and taxes*:-Services provided by e-governance in this area's:-

- Online Transaction
- Payment of Bill
- Payment of taxes
- Payment of house EMIs

Various Projects:-

1. FRIENDS: - This project is started by Kerala Government for its citizens to make online payment of electricity and water bills, revenue taxes, license fees, motor vehicle taxes, university fees, etc [1].
2. E-SEVA:-Electronic seva by Andhra Pradesh government to pay utility bills, avail of tradelicenses and transact on government matters at these facilities.
3. BWSSB ganakeekruthaGrahakaraSeve, water billing, and collection system: - This e-governance project is started by the Bangalore government. In this every month bills of houses are generated through BGS software
4. DOMESTIC: - This project is started by Daman and Diu. It is an Electricity Billing System for domestic consumers.
5. E-Pourasabha Municipal Application:-E-Pourasabha is an e-governance application for urban local bodies. It is implemented for Tax Collection system, Property Tax, Water Tax etc.
6. E-Mitra by the Government of Rajasthan
7. SAMPARK by Chandigarh government
8. E-Suvidha by the government of Uttar Pradesh

*2.1.3 Information and public relation key services:-*With these kinds of projects people can get any kind of information with just a single click.

Various projects:-

1. LokMitra: - By the government of Himachal Pradesh. The services offered include information about vacancies, tenders, market rates, matrimonial services, village e-mail.
2. Mahiti Shakti: - By Gujarat government to provide information related to its working to its citizens.
3. OLTP: - By Andhra Pradesh government. With this project 16 government departments in Andhra Pradesh are connected on a single network.

*2.1.4Municipal services*: - Services provided are as:-

- House Tax Assessment, Billing and Collection.
- Maintain records of Land & property.
- Issue of Death Certificates.
- Registration & Attorneys of properties.
- Review and approval authority for site plans

Various projects:-

1. E-Panjeeyan:-It is started by Assam government to deals with the computerization of the Document registration work at Sub Registrar Office.
2. SDO Suite:-By Assam government. This system helps in issuing various certificates like Land sale Permission, Legal heir certificate, Issue of Passport Verification Certificate, Birth and Death Report,
3. Palike:-The Palike-property tax software capture the basic details of the owner and property, payment details for which receipt is generated and given to the citizen was hosted.
4. Rural Digital Services (Nemmadi):-Provide services such as issuance of certificates, issuance of orders in respect of Social Security Schemes such as old age pension, widow pension, freedom fighter pension etc.
5. TRIS:- Tripura Registration Information System is meant for capturing of online photograph and bio-metric impression, Service for visit commission, request for duplicate document, searching of document etc.

*2.1.5Roads and traffic management:-* Services provided by this type of e-governance is:-

- Network of Roads & Bridges
- Road construction and their maintenance
- Traffic Management
- Safety ,Accident and pollution control

Various projects:-

1. BHOOSWADEENA-This project is computerized land acquisition system with tight integration with BHOOMI. The purpose of this project is to develop a system to automate the process Land acquisition
2. I-GeoApproach Internet Geometrics:-Purpose of this project is development of Geometrics based web enabled decision support system for Rural Road Network of Madhya Pradesh.
3. RSPCB (Rajasthan State Pollution Control Board);- The project relates to establishment of computer based system by fulfilling the Hardware, Software and Networking Requirements The project will be beneficial to the Government, Central Pollution Control Board, RSPCB itself.

4. CFST:- Citizen Friendly Services of Transport Department This project is started by the government of Andhra Pradesh to keeps check on pollution control, road safety, road signs and safety of its citizens.[1]

## 2.2 Areas of e-governance in rural areas:-

In rural areas e-governance has its very powerful impact. Here, from agriculture to local information everything is done through e-governance.

*2.2.1Agriculture*:- Following are the projects used in Agriculture.

1. Gyandoot: In the State of Madhya Pradesh it is an Intranet-based Government to citizen (G2c) service delivery initiative.
2. BELE:- It is a web-based application with 3-tier architecture for capturing and monitoring the major activities and services.
3. AGMARKNET: - It is a project approved by Department of Marketing & Inspection (DMI), Ministry of Agriculture, and Government of India.
4. SEEDNET:-It is a SEED informatics network under ministry of Agriculture, Government of India. The project was started in Chhattisgarh in the month of July' 2008 for Kharif season.
5. Mustard Procurement Management System:- It is started by Haryana government. It Conducted the Survey of mustard sown by the farmers and feed this data in to the database of the system. This data is then processed and generate coupons having information of dates on which farmer may visit in the mandi to sale his mustard.

*2.2.2 Local information*: - For local information such as prices of seeds, fertilizers, loan rates etc. government has started e-governance service in this area also.

Various projects:-

1. E-JanSampark:-Services & Information accessible to the common man in his locality to meet his basic need. This project is started by Chandigarh.
2. Prajavani: - it is started by the Government of Andhra Pradesh.it is a Web based On-line Monitoring of Public Grievances.
3. WebPortalsforHyderabadandCyberabadPolice:-It is designed by hyderabad, developed and hosted with many exciting public utility features like Safety tips for all citizens, verificatiosn status of Passports, Stolen vehicles etc.[1]
4. Intranet Portal of Chandigarh Administration:-It provides an environment where administration could interact.
5. E-DISHA EkalSewa Kendra:-This project is started by Haryana government.E-Disha to deliver any service from any counter/location, so at the peak requirements of services, counters can be extended as per crowd.
6. E-Samadhan: - the Government of Himachal Pradesh stressed upon to develop grievances redressed mechanism so that the genuine public grievances may be redressed in a time bound manner.

*2.2.3 Disaster management:* - To manage disaster, is a very big challenge for the government as these are natural phenomena and are unpredictable. To deal with these disasters, much state government has started e-governance service for this.

Project in this area is:-

1. Chetana: - It is a Disaster management system which has been started in the state of Bihar to deal with natural disasters such as flood and earthquake.

*2.2.4 Land record management*:-.By facilatinge-governance service in this area, millions of land records can be maintain in a very short time span.

Major projects in this area are:-

1. Bhoomi:-It is the first e-Governance land records management system project which is successfully implemented for the benefits of the common man by the Government of Karnataka.
2. Comprehensive Modernization of Land Records (CMLR):- This project is started by the government of Andhra Pradesh. It allows integrating functions of property registration, mutations and updating of field survey maps.
3. Land Record Computerisation: - The objective of the project is to computerize fresh allotment, land transfer, regularisation of occupied land etc. related actives of the Dept. of Land Management at district level.
4. Gyandoot: - it is an intranet in Dhar district of Madhya Pradesh, connecting rural cybercafés catering to the everyday needs of the masses [4].
5. Land Records Management System State Government of Punjab.
6. Devbhoomi State Government of Uttarakhand.
7. Bhu-Lekh – UP State Government of Uttar Pradesh.
8. E-Dhara State Government of Gujarat.

*2.2.5 Panchayat:-* Services provided by e-governance in this area are:-

- Issue of Birth/Death certificate.
- Application for inclusion of name in Voter list.
- Conducting various welfare schemes for the poor and needy sections of the society.
- Preparing district wise planning, implementing those plan, and review for success.
- To provide wage employment to the needy from amongst the poorest section of the rural society.
- Rural water supply and sanitation.

Various projects:-

1. E-GramViswa Gram Project:- This Project Initiates e-Gram Project connecting 13716 Gram Panchayats and 6000 Citizen Common Service Centres as a part of the e-Gram connectivity Project by Gujarat.
2. RajNidhi: - "RajNidhi" is a web enabled information kiosk system developed jointly by Rajasthan state's Department of Information Technology and Rajasthan State Agency for Computer Services (Raj Comp) [4].
3. Raj-SWIFT:-The Rajasthan State's Department of Information Technology (DoIT) has developed Government's own Intranet called as "raj-SWIFT"[4]
4. Support for P & RD sector in Assam:-NIC, Assam State Centre has been identified as the technical consultant for e-Governance solution and Computerization of the Department of Panchayat and Rural Development.
5. SamanyaMahiti by the State Government of Karnataka

## 3 E-GOVERNANCE IN HEALTH

Service provided by these projects are as:-

- Availability of medicines
- Special health camps
- Facilities at Anganwadi canters

Various projects:-

1. Online Vaccination Appointment for International Traveller:-Citizen centric application for the purpose of vaccination of the persons proceeding abroad and issuance of International Health Certificate
2. SMS based Integrated Disease Surveillance System: - it is an SMS based Integrated Disease Surveillance System facilitates to report the occurrences of disease, number of persons affected from the area of occurrences immediately to the concerned authority.
3. Hospital OPD Appointment:-Hospital OPD Appointment System is another welfare measure undertaken by Chandigarh Administration to make life of citizens simpler.
4. NLEP (National Leprosy Eradication Program):- NLEP is web based application developed for monitoring of leprosy cases in Chhattisgarh State.
5. HEALING:-it is a Health Information system for Kerala Government which is developed and implemented for Medical Health & Family Welfare department

## 4 E-GOVERNANCE IN EDUCATION:-

Providing basic education (elementary, primary, secondary) to children

- Providing computer education to children
- Results for 10th & 12thclasses
- Information on eligibility for "Distribution of books" scheme

Various projects:-

1. CASCET:-This project is started by the Karnataka government for Education Department.
2. Online Scholarship Management System:-It is meant for the purpose of distribution of scholarships and fees reimbursement.
3. AISES (All India School Education Survey):- this project is started by Assam government. This project is used for surveying the number of schools in district. Census
4. CAPnic:-This is for the Centralized seat allotment process for professional courses and come under Kerala.
5. VHSE Examination Management System:-it has been developed to handle pre-examination related activities of the vocational higher secondary education.

## 5 CHALLENGES IN E-GOVERNANCE

In this paper we have described basically three types of challenges. These are

**Technical Challenges:-** Technical issue involve the following challenges [9] such as

  i. Interoperability
 ii. Privacy
iii. Security
 iv. Multiservice Interaction

.**Organizational Challenges:-** Organizational challenges include-

  i. Lack of Integrated Services
 ii. Lack of Key Persons
iii. Population
 iv. Different Languages

According to an officer from NIC, success factors of e-Gov projects [7] –

- o 10% Technology
- o 60% Process
- o 20% Change Management
- o Rest is luck

**Economical Challenges:-** Economical challenges [9] are as:-

  i. Cost
 ii. Maintainability
iii. Reusability
 iv. Portability

## 6 PROPOSED FUTURE TECHNOLOGY OF E-GOVERNANCE

To make E-Governance more efficient and powerful, two technologies can be boon to it. These technologies are Open Source Software and Cloud Computing. In this section, we have proposed a framework to incorporate open source and cloud computing in E-Governance.

### 6.1 Open source software

Open Source Software (OSS) is software made available in both source code and binary form, under a license which allows users to freely use, modify and redistribute the software without the need to pay royalties to the original software author.

Working of E-Governance in Open Source environment

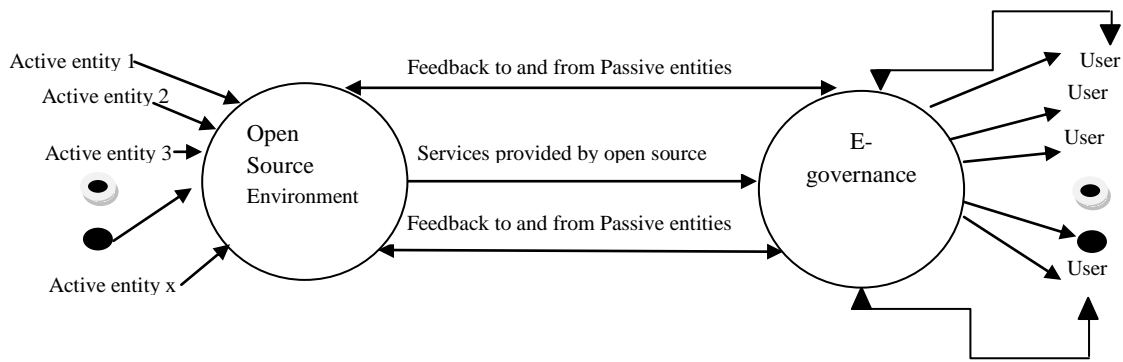

**Fig 2: E-Governance and Open Source**

Two entities that are actively involved in open source environment are active and passive entities. Active entities are code developers while passive entities are users who give their feedback openly to e-governance in open source environment. This feedback is bi-directional where an active and passive entity communicates. In this, active entities develop the code in open source environment which in turn is used by e-governance and provide its services to its citizens. Users can give their feedback o e-governance. This feedback is bidirectional.

For many governments the world over, the choice of Open Source is a strategic one. Majorities of projects which are based on Open Source technology has been highly seemed in European countries. This preference towards Open Source platforms is firstly because, acquiring and upgrading proprietary software isexpensive. There is also the proposition that it is safer to entrust knowledge in the public domain to Open Source, which is also in the public domain, than to proprietary platforms. Thirdly, using open source would enable India to encourage our own software professionals to provide software support in the form of add-on applications that could be written at a cost much lower than that required to buy multi-featured packaged software

The Indian Government has recently proposed to go the open source way in its e-governance projects, and the Department of Information Technology has issued a Draft Policy in this regard. The move has been welcomed by the IT industry. The draft titled Policy on Device Drivers for Procurement of Hardware for e-Governance states that the, "Government of India (GOI) endeavors to provide e-governance services, which are technology-neutral, cost-effective, interoperable and vendor-neutral. The GOI Policy on open standards is a step towards meeting this objective in the development of e-governance applications." The policy will be applicable in all new e-governance projects, as well as in existing ones [16].

The case for Open Source Software for governments in developing countries is a compelling one. Characteristics that make OSS an ideal candidate to be used by government are:-

1. It provides increased ownership and local autonomy.
2. It also provides increased flexibility to address localization issues and extensibility.
3. There are also numerous cost benefits to be gained from the use of OSS.

4. Many open source software projects have been extremely successful (measured by market share), particularly in server/back-office application spaces, and there is enormous industry investment in promoting the growth and improvement of community-based software projects.[8]

## 6.2 E-Governance and cloud computing

There is no doubt that worldwide revolution in Internet is changing our lives in terms of the way we work, learn and interact. Today not only private sector largely depends upon internet but government sector also depends upon internet. In this internet world, Cloud computing has become one of the hottest technique in the IT area. Today's e-governance is implemented with clouds as clouds provide number of throughout the country. NIC has selected and deployed the open source Eucalyptus software as the foundation for its cloud project, which calls for the execution of cloud-based e-governance projects on a broad scale.

NIC is providing the network backbone and a wide range of ICT (Information and Communication Technologies) services to government organizations throughout India, including a nationwide communication network for decentralized planning, improvement in government services and wider transparency of national and local government institutions [14]. Following model, we propose for usages of cloud computing in E-Governance:

Working of E-Governance with Clouds

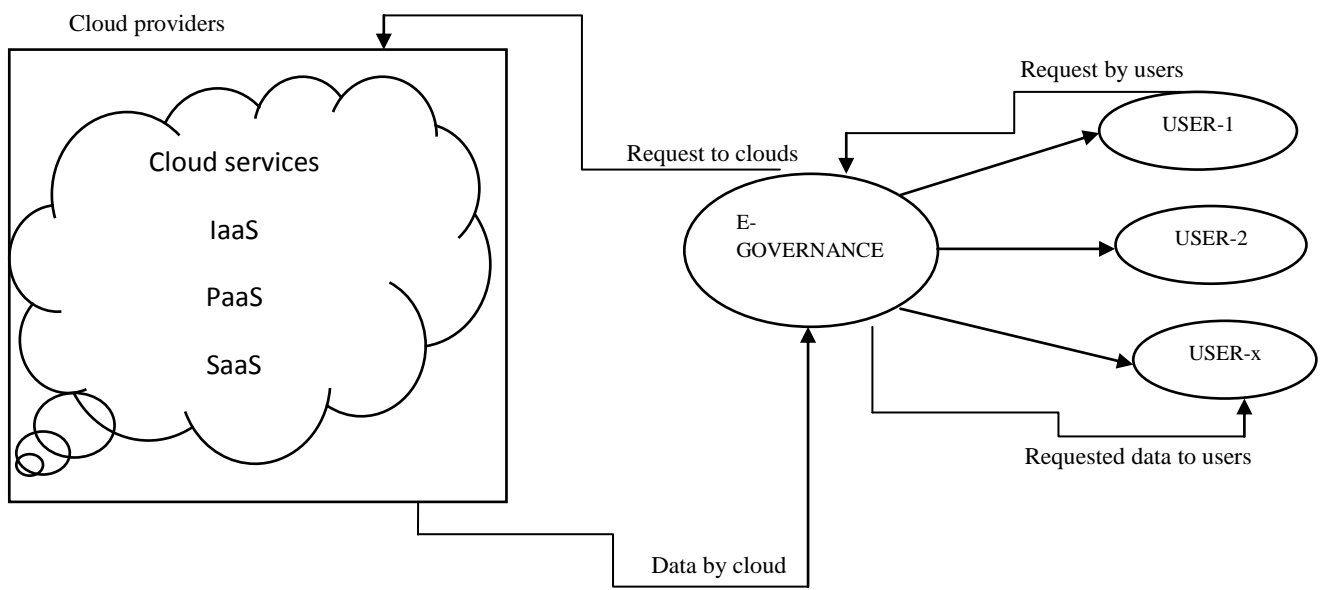

**Fig4:-E-Governance with clouds**

benefits and make e-governance more efficient to use. Cloud computing allows computer e-governance users to conveniently rent access to fully featured applications. .Cloud computing also provides software development and deployment environments, and computing infrastructure assets such as network-accessible data storage and processing model.

Cloud computing is a business model of delivering IT resources and applications as services accessible remotely over the Internet rather than locally. In the traditional, .IT resources and applications are provided in the form of products which are sold or licensed from a vendor to a user and then exploited locally on a local computer infrastructure.

SANTA BARBARA, CA-Eucalyptus Systems, creators of the most widely deployed on-premise cloud computing platform, today revealed that open source Eucalyptus Infrastructure-as-a-Service (IaaS) s/w is the cornerstone of the e-governance cloud project in India that is delivering the power of modern information technology (IT) to remote districts and villages

Clouds provide services (IaaS, PaaS and SaaS means Infrastructure as a service, Platform as a service and Software as a service) which in turn are consumed by e-governance. These services are used by e-governance and in turn provide services to its number of customers at the same time. With clouds, when loads increase a lot even then its performance doesn't decrease.

Now a days, cloud provider use Storage Area Network (SAN) to improve the storage quality of clouds. SAN provide large scale storage infrastructure to cloud provider so that they can accessed their management and evolution of heterogeneous SAN environment efficiently [18].

### 6.2.1 Need of clouds for E-Governance

Implementing e-governance without clouds is an old technique and is not too much efficient. There are number of reason which make cloud computing hottest Buzzword not in IT area but also in e-governance implementation

1. Data Integrity:-To deal with large data over the years for E-Governance applications, data should be integrals and scalable at higher level.
2. Security Auditing: - As corruption is increasing in government sector also, .somemechanism is needed to keep track on e-governance worker so that no changes in personal data of individual which is on.

In below figure, various services provided by clouds to e-governance has been shown. Almost every kinf of service is provided by clouds from designing to testing, storage to hosting and networking to security.

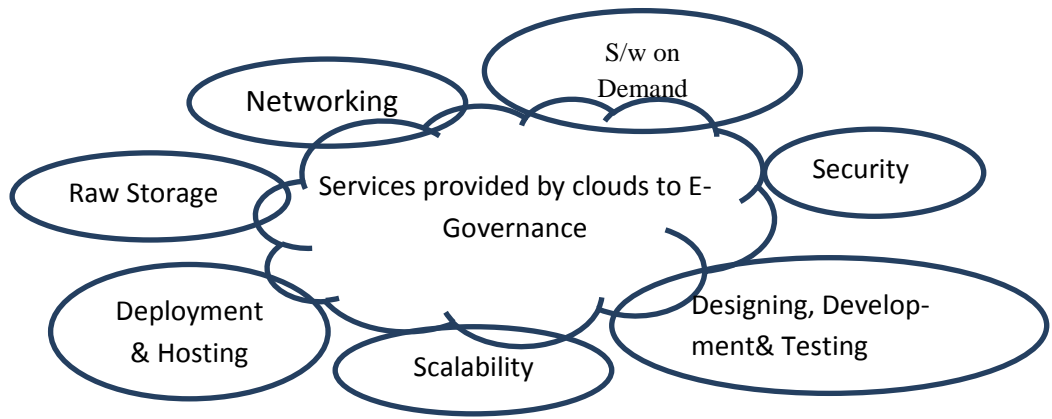

**Fig3: Cloud services**

3. Data recovery: - clouds provides an excellent way of data restoring facility which might be lost due to floods, earthquake or from any other disaster.
4. Performance and Efficiency: - An efficient e-governance is that in which lakhs of access can be done easily without decreasing the performance.
5. Systems Integration and Legacy Software:-Not only are the applications that are already providing services to be moved to the cloud, but also integrate with applications deployed in the cloud. Cloud is built on SOA principles and can offer excellent solutions for integration of various applications [10].
6. Migration to New Technologies:-Technology migration is the biggest challenge.Moving todifferent versions of software, applying application and security patches is the key to maintaining a secure data center for E-Governance. Cloud architecture efficiently enables these kinds of requirements, by co-existing and co-locating different versions and releases of the software at the same time [11].
7. Environment Protection:-Air conditioner, power usage, heat generated by powerful machines and electronic wastage by different data centered up to the pollution. This pollution can be minimized to much extend by having centralized database inspire of distributed. This is all what cloud computing do in going green.

### 6.2.2 Benefits of cloud computing

- On-demand self-sufficient service
- network access
- Location independent resource
- Rapid elasticity

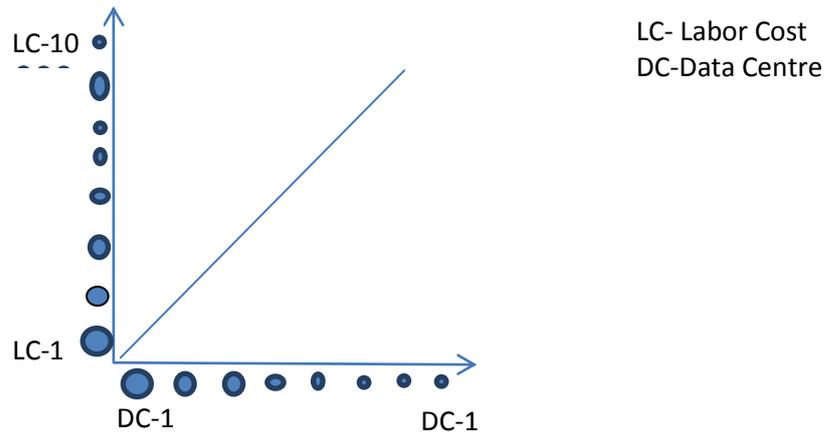

**Fig5: Data center vs. labor cost graph**

**Table 1: State wise websites for e-governance**

| CHHATTISGARH WEBSITES | DETAIL |
|---|---|
| http://agridept.cg.gov.in/ | For agriculture |
| http://siccg.gov.in/ | For any kind of information |
| http://ahd.cg.gov.in | For animal husbandry |
| http://ahd.cg.gov.in | For women and child development |
| http://www.cgdteraipur.ac.in/ | For education |
| http://nagarnigamraipur.com/ | Roads cleaning, ,propertyrecords,tax payment |
| http://www.cgsird.gov.in/ | For rural development and panchayat |
| http://comtax.cg.nic.in/comtax/ | For collecting income tax |
| http://cg.nic.in/transport/ | For transport |
| http://health.cg.gov.in/ehealth/welcome.htm | For health |
| **HARYANA** | |
| http://jamabandi.nic.in/ | For land records |

Main benefit of cloud computing is that it reduces IT laborcost [16]. In cloud computing, labor cost is almost reduces to half as compared to traditional IT approach as here database is centralized and not distributed.

| | |
|---|---|
| http://haryanaforms.nic.in/transport.asp | For transport |
| http://haryanaforms.nic.in/bd.asp | For birth and death registration |
| http://haryanaforms.nic.in/revenue.asp | For caste, ,land,income and rural certificate |
| http://haryanaforms.nic.in/ph.asp | For sewage and water connection |
| http://haryana.gov.in/employee/emp.asp | For employee |
| **CHANDIGARH** | |
| http://chandigarh.nic.in/how_know-rj.htm | For any kind of legal issue |
| http://chandigarh.nic.in/how_know-prop.htm | For property details |
| http://mcchandigarh.gov.in/ | For municipality |
| http://chdeducation.gov.in/ | For education |
| http://chandigarh.nic.in/dept_agri.htm | For agriculture |
| http://chandigarh.nic.in/dept_hlt.htm | For health |
| http://sampark.chd.nic.in/pls/esampark_web/ | For any kind of bill payments |
| **GUJRAT** | |
| http://www.gujaratinformatics.com/ | For any kind of information |
| http://www.egovamc.com/ | For land,tax,property,roads |
| **DAMAN AND DIU** | |
| http://daman.nic.in/flood_control/index.asp | For flood disaster |
| daman.nic.in/citizen_charters/agriculture.pdf | For agriculture |
| http://daman.nic.in/citizen_charters/cdpo.pdf | For child development |
| http://daman.nic.in/citizen_charters/education.pdf | For education |
| http://http://daman.nic.in/citizen_charters/directorate-Medical-&-Health-citizen-charter.pdf | For health |
| http://daman.nic.in/citizen_charters/2012/transport-citizen-charter.pdf | For transport |
| **HIMACHAL PRADESH** | |
| http://causelists.nic.in/hp/index1.html | For legal issues |
| http://admis.hp.nic.in/himpol/Citizen/OnlineComplaints.aspx | For police complaints |
| http://hpseb.com/mybill/ | For payment of electricity bills |
| http://hrtc.gov.in/hrtctickets/ | For transport |
| http://himachal.gov.in/jobportal/ | For employment |

| **ORISSA** | |
|---|---|
| http://www.agriorissa.org/ | For agriculture |
| http://www.orissa.gov.in/RD/index.htm | For rural development |
| http://www.orissa.gov.in/revenue/index.htm | Revenue and disaster mgmt. |
| http://www.orissa.gov.in/health_portal/index.html | For health |
| http://www.orissa.gov.in/commerce&transport/ | For transport |
| **TAMIL NADU** | |
| http://www.tn.gov.in/services/services_default.html#Agriculture | For agriculture |
| http://www.tn.gov.in/services/services_default.html#Health | For health |
| http://www.tn.gov.in/services/services_default.html#Higher | For education |
| http://www.tn.gov.in/services/services_default.html#Housing | For urban and rural development |
| **UTTAR PRADESH** | |
| http://gis.up.nic.in:8080/srishti/ | For district information |
| http://bhulekh.up.nic.in/ | For land records |
| http://koshvani.up.nic.in/ | Provides state financial health |
| http://rahat.up.nic.in/ | For disaster management |

## 7 CONCLUSION

In this paper, we have given a framework and application of E-Governance along with a list E-Governance projects run by state and central governments. We have also proposed future technology for E-Governance with pictorial representation of working of E-Governance with new technology. We have also proposed benefits of clouds with a graph showing how clouds reduce labor cost. Implementing E-Governance without cloud computing and open source is an old technology. Cloud computing and open source is a hottest buzzword in IT sector and we should make best possible use of these emerging technology. There are number of reasons which make cloud and open source technology so famous in E-Governance. These technologies not only provide organization, technical benefits but also provide economical benefits. E-Governance with open source is very popular in west countries but in India it is still an emerging technology. NIC is providing the network backbone and a wide range of ICT (Information and Communication Technologies) services to government organizations throughout India. Several drafts have also been passed by IT department to implement E-Governance with these technologies so that working, efficiency, transparency and security can be increased in E-Governance